\documentstyle[prb,aps]{revtex}
\begin{document}

\title{Corrections to the RPA Dielectric Function Induced by Fluctuations
 in the Momentum Distribution}

\author{Girish S. Setlur and Yia-Chung Chang}
\address{Department of Physics and Materials Research Laboratory,\\
 University of Illinois at Urbana-Champaign , Urbana Il 61801}
\maketitle

\begin{abstract}
 In this article, we introduce a new dielectric function that is different
 even from the generalised RPA, first introduced in 
 cond-mat/9810043, in that it pays attention to fluctuations in the momentum
 distribution. We argued that this is important when the momentum
 distribution is very nonideal. This new dielectric function reduces to
 the generalised RPA when the number-fluctuations are small, and
 the generalised RPA in turn reduces to the traditional RPA
 dielectric function when the momentum distribution is close to ideal.
\end{abstract}

\hspace{13in}

Let us write the generalised RPA hamiltonian and try and compute the
 dielectric function.
\begin{equation}
H_{0} = \sum_{ {\bf{k}} } \epsilon_{ {\bf{k}} }n_{0}( {\bf{k}} )
 -
\sum_{ {\bf{q}} \neq 0 }
\frac{ v_{ {\bf{q}} } }{2V}
 \sum_{ {\bf{k}} }n_{0}( {\bf{k}}+{\bf{q}}/2 )n_{0}( {\bf{k}}-{\bf{q}}/2 )
\end{equation}
\begin{equation}
H_{ext}(t) = \sum_{ {\bf{q}} }(U_{ext}({\bf{q}}t) + U^{*}_{ext}(-{\bf{q}}t))
\sum_{ {\bf{k}} }n_{ {\bf{q}} }({\bf{k}})
\end{equation}
here $ n_{ {\bf{q}} }({\bf{k}}) = c^{\dagger}_{ {\bf{k}} + {\bf{q}}/2 } c_{  {\bf{k}} - {\bf{q}}/2 } $
and $ n_{ 0 }({\bf{k}}) =  c^{\dagger}_{ {\bf{k}} }c_{ {\bf{k}} } $
Let us now write down the quation of motion for $ n_{ {\bf{q}} }({\bf{k}}) $.
\[
i\frac{ \partial }{ \partial t}n_{ {\bf{q}} }({\bf{k}}) 
 = \sum_{ {\bf{k}}^{'} }\epsilon_{ {\bf{k}}^{'} }
[n_{ {\bf{q}} }({\bf{k}}), n_{0}( {\bf{k}}^{'} )]
 - \sum_{ {\bf{q}}^{'} \neq 0 }
\frac{ v_{ {\bf{q}}^{'} } }{V}
 \sum_{ {\bf{k}}^{'} }
[n_{ {\bf{q}} }({\bf{k}}),n_{0}( {\bf{k}}^{'})]
n_{0}( {\bf{k}}^{'}-{\bf{q}}^{'} )
\]
\begin{equation}
+ \sum_{ {\bf{k}}^{'},{\bf{q}}^{'} }
(U_{ext}({\bf{q}}^{'}t) + U^{*}_{ext}(-{\bf{q}}^{'}t))
[n_{ {\bf{q}} }({\bf{k}}),n_{ {\bf{q}}^{'} }( {\bf{k}}^{'})]
\end{equation}
Now,
\[
[n_{ {\bf{q}} }({\bf{k}}),n_{ {\bf{q}}^{'} }( {\bf{k}}^{'})]
 = [c^{\dagger}_{ {\bf{k}} + {\bf{q}}/2 }c_{ {\bf{k}} - {\bf{q}}/2 },
 c^{\dagger}_{ {\bf{k}}^{'} + {\bf{q}}^{'}/2 }
c_{ {\bf{k}}^{'} - {\bf{q}}^{'}/2 }]
\]
\begin{equation}
 = c^{\dagger}_{ {\bf{k}} + {\bf{q}}/2 }c_{ {\bf{k}}^{'} - {\bf{q}}^{'}/2 }
\delta_{ {\bf{k}} - {\bf{q}}/2, {\bf{k}}^{'} + {\bf{q}}^{'}/2 }
 - c^{\dagger}_{ {\bf{k}}^{'} + {\bf{q}}^{'}/2 }c_{ {\bf{k}} - {\bf{q}}/2 }
\delta_{ {\bf{k}} + {\bf{q}}/2, {\bf{k}}^{'} - {\bf{q}}^{'}/2 }
\end{equation}
One may approximate this as,
\[
[n_{ {\bf{q}} }({\bf{k}}),n_{ {\bf{q}}^{'} }( {\bf{k}}^{'})]
 = [c^{\dagger}_{ {\bf{k}} + {\bf{q}}/2 }c_{ {\bf{k}} - {\bf{q}}/2 },
 c^{\dagger}_{ {\bf{k}}^{'} + {\bf{q}}^{'}/2 }
c_{ {\bf{k}}^{'} - {\bf{q}}^{'}/2 }]
\]
\begin{equation}
 = [ n_{0}({\bf{k}}+{\bf{q}}/2) - n_{0}({\bf{k}}-{\bf{q}}/2) ]
\delta_{ {\bf{k}}, {\bf{k}}^{'} }\delta_{ {\bf{q}}, -{\bf{q}}^{'} }
\end{equation}
 The next approximation would be to replace the number operator by its c-number
 expectation value. But we shall desist from that for the moment.
\begin{equation}
[n_{ {\bf{q}} }({\bf{k}}), n_{0}( {\bf{k}}^{'} )]
 = n_{ {\bf{q}} }({\bf{k}})(\delta_{ {\bf{k}}^{'} , {\bf{k}}-{\bf{q}}/2 }
 - \delta_{ {\bf{k}}^{'} , {\bf{k}}+{\bf{q}}/2 })
\end{equation}
\[
i\frac{ \partial }{ \partial t}n_{ {\bf{q}} }({\bf{k}}) 
 = (\epsilon_{ {\bf{k}}-{\bf{q}}/2 }-\epsilon_{ {\bf{k}}+{\bf{q}}/2 })
 n_{ {\bf{q}} }({\bf{k}})
 - \sum_{ {\bf{q}}^{'} \neq 0 }
\frac{ v_{ {\bf{q}}^{'} } }{V}
(n_{0}( {\bf{k}}-{\bf{q}}/2-{\bf{q}}^{'} )
- n_{0}({\bf{k}}+{\bf{q}}/2-{\bf{q}}^{'} ))
 n_{ {\bf{q}} }({\bf{k}})
\]
\begin{equation}
+ 
(U_{ext}(-{\bf{q}}t) + U^{*}_{ext}({\bf{q}}t))
(n_{0}({\bf{k}}+{\bf{q}}/2) - n_{0}({\bf{k}}-{\bf{q}}/2))
\end{equation}
\newpage
\begin{equation}
\langle n_{ {\bf{q}} }({\bf{k}}) \rangle
 = U_{ext}(-{\bf{q}}t)C_{ {\bf{q}} }({\bf{k}})
+  U^{*}_{ext}({\bf{q}}t)D_{ {\bf{q}} }({\bf{k}})
\end{equation}
Let us now ignore fluctuations in the momentum distribution. That is,
 we are allowed to replace
\begin{equation}
\langle n_{ 0 }({\bf{k}}^{'}) n_{ {\bf{q}} }({\bf{k}}) \rangle
 = \langle n_{ 0 }({\bf{k}}^{'}) \rangle\langle n_{ {\bf{q}} }({\bf{k}}) \rangle
\end{equation}
\[
\omega\mbox{      } U_{ext}(-{\bf{q}}t)C_{ {\bf{q}} }({\bf{k}})
- \omega\mbox{      }U^{*}_{ext}({\bf{q}}t)D_{ {\bf{q}} }({\bf{k}})
 =  (\epsilon_{ {\bf{k}}-{\bf{q}}/2 }-\epsilon_{ {\bf{k}}+{\bf{q}}/2 })
 U_{ext}(-{\bf{q}}t)C_{ {\bf{q}} }({\bf{k}})
+  (\epsilon_{ {\bf{k}}-{\bf{q}}/2 }-\epsilon_{ {\bf{k}}+{\bf{q}}/2 })
 U^{*}_{ext}({\bf{q}}t)D_{ {\bf{q}} }({\bf{k}})
\]
\[
 - \sum_{ {\bf{q}}^{'} \neq 0 }
\frac{ v_{ {\bf{q}}^{'} } }{V}
(\langle n_{0}( {\bf{k}}-{\bf{q}}/2-{\bf{q}}^{'} ) \rangle
- \langle n_{0}({\bf{k}}+{\bf{q}}/2-{\bf{q}}^{'} ) \rangle )
  U_{ext}(-{\bf{q}}t)C_{ {\bf{q}} }({\bf{k}})
\]
\[
- \sum_{ {\bf{q}}^{'} \neq 0 }
\frac{ v_{ {\bf{q}}^{'} } }{V}
(\langle n_{0}( {\bf{k}}-{\bf{q}}/2-{\bf{q}}^{'} ) \rangle
- \langle n_{0}({\bf{k}}+{\bf{q}}/2-{\bf{q}}^{'} ) \rangle )
  U^{*}_{ext}({\bf{q}}t)D_{ {\bf{q}} }({\bf{k}})
\]
\begin{equation}
+ 
(U_{ext}(-{\bf{q}}t) + U^{*}_{ext}({\bf{q}}t))
(\langle n_{0}({\bf{k}}+{\bf{q}}/2) \rangle 
 - \langle n_{0}({\bf{k}}-{\bf{q}}/2) \rangle)
\end{equation}
\[
\omega\mbox{      }C_{ {\bf{q}} }({\bf{k}})
 = (\epsilon_{ {\bf{k}}-{\bf{q}}/2 }-\epsilon_{ {\bf{k}}+{\bf{q}}/2 })
C_{ {\bf{q}} }({\bf{k}})- \sum_{ {\bf{q}}^{'} \neq 0 }
\frac{ v_{ {\bf{q}}^{'} } }{V}
(\langle n_{0}( {\bf{k}}-{\bf{q}}/2-{\bf{q}}^{'} ) \rangle
- \langle n_{0}({\bf{k}}+{\bf{q}}/2-{\bf{q}}^{'} ) \rangle )
 C_{ {\bf{q}} }({\bf{k}})
\]
\begin{equation}
+ 
(\langle n_{0}({\bf{k}}+{\bf{q}}/2) \rangle
 - \langle n_{0}({\bf{k}}-{\bf{q}}/2) \rangle)
\end{equation}
Since,
\begin{equation}
U_{eff}({\bf{q}}t) = U_{ext}({\bf{q}}t) + \frac{ v_{ {\bf{q}} } }{V}
\langle \rho^{'}_{ -{\bf{q}} } \rangle U_{ext}({\bf{q}}t) 
\end{equation}
\begin{equation}
\langle \rho^{'}_{ -{\bf{q}} } \rangle
 = \sum_{ {\bf{k}} }C_{ -{\bf{q}} }({\bf{k}})
\end{equation}
\begin{equation}
 \sum_{ {\bf{k}} }C_{ -{\bf{q}} }({\bf{k}})
 = \sum_{ {\bf{k}} }
\frac{ \langle n_{0}({\bf{k}}-{\bf{q}}/2) \rangle
 - \langle n_{0}({\bf{k}}+{\bf{q}}/2) \rangle }
{\omega - {\tilde{\epsilon}}_{ {\bf{k}} + {\bf{q}}/2 }
 + {\tilde{\epsilon}}_{ {\bf{k}} - {\bf{q}}/2 } }
\end{equation}
If we use the definition of the dielectric function we get precisely the
 WRONG ANSWER !!!
\begin{equation}
\epsilon_{WRONG}({\bf{q}}, \omega) = 
\frac{ U_{ext}({\bf{q}}t) }{  U_{eff}({\bf{q}}t) }
 = \frac{1}{ 1 + \frac{ v_{ {\bf{q}} } }{V}
\sum_{ {\bf{k}} }
\frac{ \langle n_{0}({\bf{k}}-{\bf{q}}/2) \rangle
 - \langle n_{0}({\bf{k}}+{\bf{q}}/2) \rangle }
{\omega - {\tilde{\epsilon}}_{ {\bf{k}} + {\bf{q}}/2 }
 + {\tilde{\epsilon}}_{ {\bf{k}} - {\bf{q}}/2 } } }
\end{equation}
 The reason is because we have been very cavalier in our treatment of
 correlations. Parenthetically we note that when I said this is the wrong
 answer, I meant it is not the RPA dielectric function. But then so what ??
 Nobody told me RPA was a controlled approximation. But then we must press
 on, and try to find a clever way of reproducing the RPA dielectric function.
 One has to consider the full Hamiltonian when dealing
 with the dielectric function rather than just $ H_{0} $.
 Let us now try and do this.
\[
i\frac{ \partial }{ \partial t} n^{t}_{ {\bf{q}} }({\bf{k}})
 = (\epsilon_{ {\bf{k}}-{\bf{q}}/2 } - \epsilon_{ {\bf{k}}+{\bf{q}}/2 })
 n^{t}_{ {\bf{q}} }({\bf{k}}) 
+ \sum_{ {\bf{q}}^{'} \neq 0 }\frac{ v_{ {\bf{q}}^{'} } }{2V}
[n_{ {\bf{q}} }({\bf{k}}), \rho_{ {\bf{q}}^{'} }] \rho^{t}_{ -{\bf{q}}^{'} }
+ \sum_{ {\bf{q}}^{'} \neq 0 }\frac{ v_{ {\bf{q}}^{'} } }{2V}
\rho^{t}_{ {\bf{q}}^{'} }[n_{ {\bf{q}} }({\bf{k}}), \rho_{ -{\bf{q}}^{'} }] 
\]
\begin{equation}
 + \sum_{ {\bf{q}}^{'} \neq 0 }(U_{ext}({\bf{q}}^{'}t) 
+ U^{*}_{ext}(-{\bf{q}}^{'}t))
[n_{ {\bf{q}} }({\bf{k}}),\rho_{ {\bf{q}}^{'} }]
\end{equation}
\[
[n_{ {\bf{q}} }({\bf{k}}),\rho_{ {\bf{q}}^{'} }]
 = \sum_{ {\bf{k}}^{'} }
[c^{\dagger}_{ {\bf{k}} + {\bf{q}}/2 } c_{  {\bf{k}} - {\bf{q}}/2 },
c^{\dagger}_{ {\bf{k}}^{'} + {\bf{q}}^{'}/2 }
 c_{  {\bf{k}}^{'} - {\bf{q}}^{'}/2 }]
\]
\[
 = \sum_{ {\bf{k}}^{'} }
c^{\dagger}_{ {\bf{k}} + {\bf{q}}/2 }c_{  {\bf{k}}^{'} - {\bf{q}}^{'}/2 }
\delta_{ {\bf{k}} - {\bf{q}}/2, {\bf{k}}^{'} + {\bf{q}}^{'}/2 }
 - \sum_{ {\bf{k}}^{'} }
c^{\dagger}_{ {\bf{k}}^{'} + {\bf{q}}^{'}/2 }c_{  {\bf{k}} - {\bf{q}}/2 }
\delta_{ {\bf{k}} + {\bf{q}}/2, {\bf{k}}^{'} - {\bf{q}}^{'}/2 }
\]
\begin{equation}
\approx \delta_{ {\bf{q}}, -{\bf{q}}^{'} }
[n_{0}({\bf{k}} + {\bf{q}}/2) - n_{0}({\bf{k}} - {\bf{q}}/2)]
\end{equation}
\[
i\frac{ \partial }{ \partial t} n^{t}_{ {\bf{q}} }({\bf{k}})
 = (\epsilon_{ {\bf{k}}-{\bf{q}}/2 } - \epsilon_{ {\bf{k}}+{\bf{q}}/2 })
 n^{t}_{ {\bf{q}} }({\bf{k}}) 
+ \frac{ v_{ {\bf{q}} } }{V}
(n_{0}({\bf{k}} + {\bf{q}}/2) - n_{0}({\bf{k}} - {\bf{q}}/2))
 \rho^{t}_{ {\bf{q}} }
\]
\begin{equation}
 + (U_{ext}(-{\bf{q}}t) 
+ U^{*}_{ext}({\bf{q}}t))
(n_{0}({\bf{k}} + {\bf{q}}/2) - n_{0}({\bf{k}} - {\bf{q}}/2))
\end{equation}
Let us make a first pass at the computation of the dielectric function.
Here, we make use of mean-field theory, that is, replace
 $ \langle n_{0}( {\bf{k}}^{'} )\rho_{ {\bf{q}} } \rangle = \langle n_{0}( {\bf{k}}^{'} ) \rangle\langle \rho_{ {\bf{q}} } \rangle $
\[
\omega \mbox{      }\langle  n^{t}_{ {\bf{q}} }({\bf{k}}) \rangle
 = (\epsilon_{ {\bf{k}}-{\bf{q}}/2 } - \epsilon_{ {\bf{k}}+{\bf{q}}/2 })
\langle  n^{t}_{ {\bf{q}} }({\bf{k}}) \rangle
+ \frac{ v_{ {\bf{q}} } }{V}
(\langle  n_{0}({\bf{k}} + {\bf{q}}/2)  \rangle
 -  \langle  n_{0}({\bf{k}} - {\bf{q}}/2)  \rangle)
\langle  \rho^{t}_{ {\bf{q}} } \rangle
\]
\begin{equation}
 + (U_{ext}(-{\bf{q}}t)
+ U^{*}_{ext}({\bf{q}}t))
(\langle  n_{0}({\bf{k}} + {\bf{q}}/2) \rangle -
\langle  n_{0}({\bf{k}} - {\bf{q}}/2) \rangle)
\end{equation}
\[
\langle  n^{t}_{ {\bf{q}} }({\bf{k}}) \rangle
 = \frac{ v_{ {\bf{q}} } }{V}
\frac{\langle  n_{0}({\bf{k}} + {\bf{q}}/2)  \rangle
 -  \langle  n_{0}({\bf{k}} - {\bf{q}}/2)  \rangle }
{ \omega  - \epsilon_{ {\bf{k}}-{\bf{q}}/2 } +
 \epsilon_{ {\bf{k}}+{\bf{q}}/2 }  }\langle  \rho^{t}_{ {\bf{q}} } \rangle
\]
\begin{equation}
 + (U_{ext}(-{\bf{q}}t)
+ U^{*}_{ext}({\bf{q}}t))
\frac{\langle  n_{0}({\bf{k}} + {\bf{q}}/2)  \rangle
 -  \langle  n_{0}({\bf{k}} - {\bf{q}}/2)  \rangle }
{ \omega  - \epsilon_{ {\bf{k}}-{\bf{q}}/2 }  +\epsilon_{ {\bf{k}}+{\bf{q}}/2 }  }
\end{equation}
This means,
\begin{equation}
\langle \rho_{ -{\bf{q}} } \rangle = U_{ext}({\bf{q}}t)
\frac{ P_{0}({\bf{q}},\omega) }{ \epsilon({\bf{q}}, \omega) }
\end{equation}
\begin{equation}
 P_{0}({\bf{q}},\omega) = \sum_{ {\bf{k}} }
\frac{ \langle  n_{0}({\bf{k}} - {\bf{q}}/2)  \rangle
 -  \langle  n_{0}({\bf{k}} + {\bf{q}}/2)  \rangle }
{\omega - \epsilon_{ {\bf{k}} + {\bf{q}}/2 } 
+ \epsilon_{ {\bf{k}} - {\bf{q}}/2 } }
\end{equation}
\begin{equation}
\epsilon({\bf{q}}, \omega) = 1 - \frac{v_{ {\bf{q}} }}{V} 
P_{0}({\bf{q}},\omega)
\end{equation}
From this and the fact that 
\begin{equation}
\epsilon_{g-RPA}({\bf{q}},\omega) = \frac{ U_{ext}({\bf{q}}t) }
{  U_{eff}({\bf{q}}t) } = \epsilon({\bf{q}},\omega) 
\end{equation}
Next we would like to include fluctuations. Let us do this differently
 this time via the use of the BBGKY heirarchy.
\[
i\frac{ \partial }{ \partial t} \langle n^{t}_{ {\bf{q}} }({\bf{k}}) \rangle
 = (\epsilon_{ {\bf{k}}-{\bf{q}}/2 } - \epsilon_{ {\bf{k}}+{\bf{q}}/2 })
 \langle n^{t}_{ {\bf{q}} }({\bf{k}})  \rangle
+ \frac{ v_{ {\bf{q}} } }{V}
(\langle n_{0}({\bf{k}} + {\bf{q}}/2)  \rangle -
 \langle n_{0}({\bf{k}} - {\bf{q}}/2)  \rangle)
\langle  \rho^{t}_{ {\bf{q}} } \rangle
\]
\[
+  \frac{ v_{ {\bf{q}} } }{V}
(F_{2A}({\bf{k}} + {\bf{q}}/2, {\bf{q}}) 
- F_{2A}({\bf{k}} - {\bf{q}}/2, {\bf{q}}))
\]
\begin{equation}
 + (U_{ext}(-{\bf{q}}t) 
+ U^{*}_{ext}({\bf{q}}t))
(n_{0}({\bf{k}} + {\bf{q}}/2) - n_{0}({\bf{k}} - {\bf{q}}/2))
\end{equation}
Here,
\begin{equation}
F_{2A}({\bf{k}}^{'},{\bf{q}};t) = \langle n_{0}({\bf{k}}^{'})
\rho^{t}_{ {\bf{q}} } \rangle - \langle n_{0}({\bf{k}}^{'})\rangle 
 \langle \rho^{t}_{ {\bf{q}} } \rangle
\end{equation}
\begin{equation}
F_{2}({\bf{k}}^{'};{\bf{k}},{\bf{q}};t)
 = \langle n_{0}({\bf{k}}^{'})n^{t}_{ {\bf{q}} }({\bf{k}}) \rangle
 - \langle n_{0}({\bf{k}}^{'}) \rangle
 \langle n^{t}_{ {\bf{q}} }({\bf{k}}) \rangle
\end{equation}
\begin{equation}
F_{2A}({\bf{k}}^{'},{\bf{q}};t) = \sum_{ {\bf{k}} }
F_{2}({\bf{k}}^{'};{\bf{k}},{\bf{q}};t)
\end{equation}
\[
i\frac{ \partial }{ \partial t }F_{2}({\bf{k}}^{'}; {\bf{k}},{\bf{q}};t)
 = (\epsilon_{ {\bf{k}}-{\bf{q}}/2 } - \epsilon_{ {\bf{k}}+{\bf{q}}/2 })
F_{2}({\bf{k}}^{'}; {\bf{k}},{\bf{q}};t)
 + \frac{ v_{ {\bf{q}} } }{V}(N({\bf{k}}^{'},{\bf{k}}+{\bf{q}}/2)
 - N({\bf{k}}^{'},{\bf{k}}-{\bf{q}}/2))
\langle \rho^{t}_{ {\bf{q}} } \rangle
\]
\begin{equation}
+  \frac{ v_{ {\bf{q}} } }{V}(\langle n_{0}({\bf{k}}+{\bf{q}}/2) \rangle
 - \langle n_{0}({\bf{k}}-{\bf{q}}/2) \rangle)
F_{2A}({\bf{k}}^{'},{\bf{q}};t)
 + (U_{ext}(-{\bf{q}}t) + U^{*}_{ext}({\bf{q}}t))
(N({\bf{k}}^{'}, {\bf{k}}+{\bf{q}}/2) - N({\bf{k}}^{'}, {\bf{k}}-{\bf{q}}/2))
\end{equation}
Let us write,
\begin{equation}
F_{2}({\bf{k}}^{'}; {\bf{k}},{\bf{q}};t) = U_{ext}(-{\bf{q}}t)
F_{2,a}({\bf{k}}^{'}; {\bf{k}},{\bf{q}})
 + U^{*}_{ext}({\bf{q}}t) F_{2,b}({\bf{k}}^{'}; {\bf{k}},{\bf{q}})
\end{equation}
\begin{equation}
\langle \rho^{t}_{ {\bf{q}} } \rangle = U_{ext}(-{\bf{q}}t)
\langle \rho^{'}_{ {\bf{q}} } \rangle 
+  U^{*}_{ext}({\bf{q}}t) \langle \rho^{''}_{ {\bf{q}} } \rangle
\end{equation}
Also define,
\begin{equation}
N({\bf{k}},{\bf{k}}^{'}) = \langle n_{0}({\bf{k}})n_{0}({\bf{k}}^{'}) \rangle
 - \langle n_{0}({\bf{k}}) \rangle \langle  n_{0}({\bf{k}}^{'}) \rangle
\end{equation}
\[
\omega \mbox{      }F_{2,a}({\bf{k}}^{'}; {\bf{k}},{\bf{q}})
 = (\epsilon_{ {\bf{k}}-{\bf{q}}/2 } - \epsilon_{ {\bf{k}}+{\bf{q}}/2 })
F_{2,a}({\bf{k}}^{'}; {\bf{k}},{\bf{q}})
 + \frac{ v_{ {\bf{q}} } }{V}(N({\bf{k}}^{'},{\bf{k}}+{\bf{q}}/2)
 - N({\bf{k}}^{'},{\bf{k}}-{\bf{q}}/2))
\langle \rho^{'}_{ {\bf{q}} } \rangle
\]
\begin{equation}
+  \frac{ v_{ {\bf{q}} } }{V}(\langle n_{0}({\bf{k}}+{\bf{q}}/2) \rangle
 - \langle n_{0}({\bf{k}}-{\bf{q}}/2) \rangle)
F^{a}_{2A}({\bf{k}}^{'},{\bf{q}})
 + 
(N({\bf{k}}^{'}, {\bf{k}}+{\bf{q}}/2) - N({\bf{k}}^{'}, {\bf{k}}-{\bf{q}}/2))
\end{equation}
\[
-\omega \mbox{      }F_{2,b}({\bf{k}}^{'}; {\bf{k}},{\bf{q}})
 = (\epsilon_{ {\bf{k}}-{\bf{q}}/2 } - \epsilon_{ {\bf{k}}+{\bf{q}}/2 })
F_{2,b}({\bf{k}}^{'}; {\bf{k}},{\bf{q}})
 + \frac{ v_{ {\bf{q}} } }{V}(N({\bf{k}}^{'},{\bf{k}}+{\bf{q}}/2)
 - N({\bf{k}}^{'},{\bf{k}}-{\bf{q}}/2))
\langle \rho^{''}_{ {\bf{q}} } \rangle
\]
\begin{equation}
+  \frac{ v_{ {\bf{q}} } }{V}(\langle n_{0}({\bf{k}}+{\bf{q}}/2) \rangle
 - \langle n_{0}({\bf{k}}-{\bf{q}}/2) \rangle)
F^{b}_{2A}({\bf{k}}^{'},{\bf{q}})
 + 
(N({\bf{k}}^{'}, {\bf{k}}+{\bf{q}}/2) - N({\bf{k}}^{'}, {\bf{k}}-{\bf{q}}/2))
\end{equation}
\[
\epsilon({\bf{q}},\omega) \mbox{         }F_{2A}^{a}({\bf{k}}^{'},{\bf{q}})
 = \frac{ v_{ {\bf{q}} } }{V}\sum_{ {\bf{k}} }
\frac{ N({\bf{k}}^{'}, {\bf{k}}+{\bf{q}}/2)
 -  N({\bf{k}}^{'}, {\bf{k}}-{\bf{q}}/2) }
{\omega - \epsilon_{ {\bf{k}}-{\bf{q}}/2 } +  \epsilon_{ {\bf{k}}+{\bf{q}}/2 }}
\langle \rho^{'}_{ {\bf{q}} } \rangle
\]
\begin{equation}
+ \sum_{ {\bf{k}} }\frac{ N({\bf{k}}^{'}, {\bf{k}}+{\bf{q}}/2)
 -  N({\bf{k}}^{'}, {\bf{k}}-{\bf{q}}/2) }
{\omega - \epsilon_{ {\bf{k}}-{\bf{q}}/2 } +  \epsilon_{ {\bf{k}}+{\bf{q}}/2 }}
\end{equation}
\[
\omega\mbox{   }C_{ {\bf{q}} }({\bf{k}})
 = (\epsilon_{ {\bf{k}}-{\bf{q}}/2 } - \epsilon_{ {\bf{k}}+{\bf{q}}/2 })
C_{ {\bf{q}} }({\bf{k}})
+ \frac{ v_{ {\bf{q}} } }{V}(\langle n_{0}({\bf{k}}+{\bf{q}}/2) \rangle
 - \langle n_{0}({\bf{k}}-{\bf{q}}/2) \rangle)
\langle \rho^{'}_{ {\bf{q}} } \rangle
+  \frac{ v_{ {\bf{q}} } }{V}(F^{a}_{2A}({\bf{k}}+{\bf{q}}/2) 
 - F^{a}_{2A}({\bf{k}}-{\bf{q}}/2))
\]
\begin{equation}
+ (\langle n_{0}({\bf{k}}+{\bf{q}}/2) \rangle 
- \langle n_{0}({\bf{k}}-{\bf{q}}/2) \rangle)
\end{equation}
\newpage
\[
C_{ {\bf{q}} }({\bf{k}}) = \frac{ v_{ {\bf{q}} } }{V}
\frac{ \langle n_{0}({\bf{k}}+{\bf{q}}/2) \rangle
 - \langle n_{0}({\bf{k}}-{\bf{q}}/2) \rangle }
{ \omega - \epsilon_{ {\bf{k}}-{\bf{q}}/2 } + \epsilon_{ {\bf{k}}+{\bf{q}}/2 } }
\langle \rho^{'}_{ {\bf{q}} } \rangle
\]
\[
+ \frac{ \langle n_{0}({\bf{k}}+{\bf{q}}/2) \rangle
 - \langle n_{0}({\bf{k}}-{\bf{q}}/2) \rangle }
{ \omega - \epsilon_{ {\bf{k}}-{\bf{q}}/2 } + \epsilon_{ {\bf{k}}+{\bf{q}}/2 } }
\]
\begin{equation}
+  \frac{ v_{ {\bf{q}} } }{V}
\frac{ F^{a}_{2A}({\bf{k}}+{\bf{q}}/2,{\bf{q}})
 - F^{a}_{2A}({\bf{k}}-{\bf{q}}/2,{\bf{q}}) }
{ \omega - \epsilon_{ {\bf{k}}-{\bf{q}}/2 } + \epsilon_{ {\bf{k}}+{\bf{q}}/2 } }
\end{equation}
After all this, it may be shown that the overall dielectric function including
 possible fluctutations in the momentum distribution is given by,
\begin{equation}
\epsilon_{eff}({\bf{q}},\omega) = \epsilon_{g-RPA}({\bf{q}},\omega)
 - (\frac{v_{ {\bf{q}} } }{V})^{2}\frac{ P_{2}({\bf{q}},\omega) }
{ \epsilon_{g-RPA}({\bf{q}},\omega) }
\end{equation}
Here,
\begin{equation}
P_{2}({\bf{q}},\omega) = \sum_{ {\bf{k}},{\bf{k}}^{'} }
\frac{ N({\bf{k}}+{\bf{q}}/2, {\bf{k}}^{'}+{\bf{q}}/2)
- N({\bf{k}}-{\bf{q}}/2, {\bf{k}}^{'}+{\bf{q}}/2)
- N({\bf{k}}+{\bf{q}}/2, {\bf{k}}^{'}-{\bf{q}}/2)
+ N({\bf{k}}-{\bf{q}}/2, {\bf{k}}^{'}-{\bf{q}}/2) }
{ (\omega - \epsilon_{ {\bf{k}}-{\bf{q}}/2 } +  
\epsilon_{ {\bf{k}}+{\bf{q}}/2 })
(\omega - \epsilon_{ {\bf{k}}^{'}-{\bf{q}}/2 } +
\epsilon_{ {\bf{k}}^{'}+{\bf{q}}/2 }) }
\end{equation}
\begin{equation}
\epsilon_{g-RPA}({\bf{q}},\omega) = 1 + \frac{ v_{ {\bf{q}} } }{V}
\sum_{ {\bf{k}} } \frac{ \langle n_{0}({\bf{k}}+{\bf{q}}/2) \rangle
 - \langle n_{0}({\bf{k}}-{\bf{q}}/2) \rangle }
{ \omega - \epsilon_{ {\bf{k}}+{\bf{q}}/2 } +\epsilon_{ {\bf{k}}-{\bf{q}}/2  } }
\end{equation}


\begin{thebibliography}{30}
\bibitem[8]{footnote} The pathological exception is for nonideal 
 distributions such as,
 $ \langle n_{ {\bf{k}} } \rangle = 0 $ for $ |{\bf{k}}| < 0.5\mbox{   }k_{f} $
 and $ \langle n_{ {\bf{k}} } \rangle = 1 $ for
 $ 0.5\mbox{   }k_{f} < |{\bf{k}}| < 1.5\mbox{   }k_{f} $
 and $ \langle n_{ {\bf{k}} } \rangle = 0 $ for
 $ |{\bf{k}}| > 1.5\mbox{   }k_{f} $ 
 this is very nonideal but has zero fluctuation, 
 but hopefully these are not physical.
\end{thebibliography}
\end{document}